\documentclass[%
 reprint,
showpacs,preprintnumbers,
amsmath,amssymb,
aps,
prl,
 longbibliography,
 lengthcheck%
]{revtex4-1}

\usepackage{graphicx, epsfig}
\usepackage{dcolumn}
\usepackage{bm}
\usepackage{hyperref}
\usepackage{type1cm}
\usepackage[caption=false]{subfig}

\begin{document}

\preprint{APS/123-QED}

\title{Number-resolved imaging of excited-state atoms using a scanning autoionization microscope}

\author{G. Lochead}
\author{D. Boddy}
\author{D. P. Sadler}
\author{C. S. Adams}
\author{M. P. A. Jones}
\email{m.p.a.jones@durham.ac.uk}
\affiliation{Joint Quantum Centre (JQC) Durham-Newcastle, Department of Physics, Durham University, Durham DH1 3LE, UK}

\date{\today}

\begin{abstract}
We report on a scanning microscopy technique for atom-number-resolved imaging of excited-state atoms. A tightly focussed laser beam leads to local autoionization, and the resulting ions are counted electronically. Scanning the beam across the cloud builds up an image of the density distribution of excited atoms, with access to the full counting statistics at each spatial sampling point and an overall detection efficiency of $21~\%$. We apply this technique to the measurement of a spatially inhomogeneous electric field with a spatial resolution of 50~$\mu$m and a sensitivity to electric field gradients of 0.04~V~cm$^{-2}$.
\end{abstract}

\pacs{32.80.Ee, 32.80.Rm, 032.80.Zb}
                           
\maketitle

Ultracold atomic gases constitute an almost ideal testbed for studying complex quantum phases of matter \cite{Bloch2008}. State-of-the-art experiments are now able to image the transition from a superfluid to an insulating state with single-atom resolution \cite{Sherson2010,Bakr2010}. Recently it has become clear that Rydberg states can be exploited to introduce tunable, long-range interactions, leading directly to the creation of many-body quantum states \cite{Lukin2001} that are associated with strong, long-range spatial correlations \cite{Schwarzkopf2011,Schausz2012}. With control over the geometry and excitation parameters, more exotic states are possible, where the Rydberg excitations form a crystal with true long-range order \cite{Schachenmayer2010,Pohl2010,VanBijnen2011}.

To probe these strongly correlated systems effectively, Rydberg atoms must be detected with single-atom sensitivity and micron-scale spatial resolution. Standard techniques, such as absorption and fluorescence imaging, have been applied to the detection of ions \cite{Simien2004} and Rydberg atoms \cite{McQuillen2012} in an ultracold plasma, but single-atom sensitivity is challenging as only a few photons are scattered by each atom. If the atoms are confined to individual trapping sites, single Rydberg excitations can be detected via trap loss \cite{Urban2009,Gaetan2009}. Such an approach was recently employed to measure the spatial correlation function in a strongly interacting Rydberg gas \cite{Schausz2012}. Spatial correlations have also been observed using ion imaging techniques \cite{Schwarzkopf2011}, but stray electric fields from the ion optics can lead to unwanted Stark shifts that preclude the observation of crystalline states. Other proposals include the extension of electromagnetically-induced transparency techniques \cite{Mohapatra2007,Tauchinsky2010} to detect single Rydberg excitations \cite{Olmos2011,Gunter2012}.

\begin{figure}[ht!]
  \includegraphics{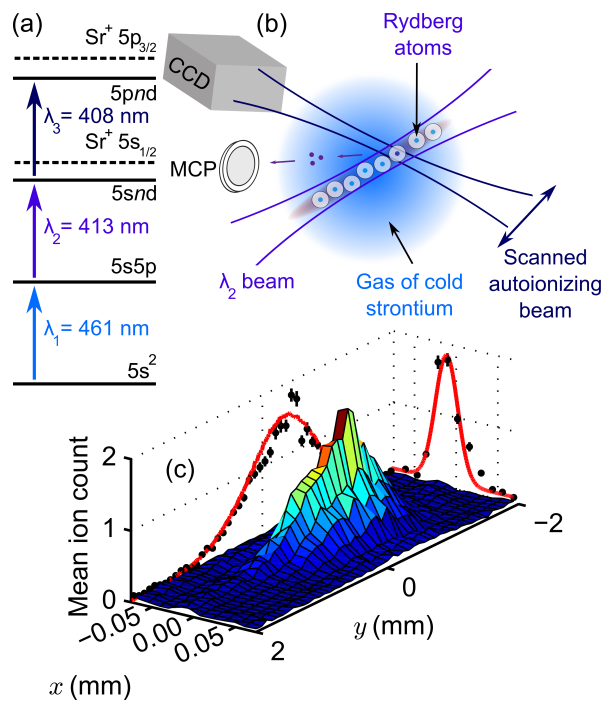}
  \caption{\label{Fig. 1} (Color online) (a) Relevant energy levels and transitions in $^{88}$Sr. (b) Overview of the experiment. The distribution of Rydberg atoms is defined by the tightly focused $\lambda_2$ beam.  (c) 2D image of the  Rydberg density distribution at $n=56$.  The $y$-axis projection shows a single slice (black points). The solid red line is the corresponding slice through the ground-state fluorescence image, with the amplitude scaled to give the best overlap.  The  $x$-axis projection shows a slice through the centre of the distribution overlaid with a Gaussian fit (red solid line).}
  \label{fig:setup}
\end{figure}

Here we report on a novel technique for the measurement of the spatial distribution of excited-state atoms, based upon core excitation \cite{Cooke1978} of a two-electron atom. If one valence electron is in an excited state (typically principal quantum number $n  > 10$), excitation of the second valence electron leads to rapid and highly efficient autoionization.  Using a tightly focused autoionization laser, we selectively ionize the Rydberg atoms in a small spatial region. The resulting ions are counted using an electronic detector (see Fig.~\ref{Fig. 1}), with an overall efficiency of up to 20~\%. By scanning the laser across the cloud an image of the spatial distribution of the Rydberg atoms is built up, with single-atom sensitivity and full counting statistics at each position. Crucially, this technique operates independently of any external fields, such as trap potentials or electric fields. We apply this technique to the spatially resolved measurement of an inhomogeneous electric field, achieving a sensitivity to electric field gradients of 0.04~V~cm$^{-2}$ with a spatial resolution of 50~$\mu$m.


The extracted electric field compares well to other Rydberg static electrometry techniques \cite{Tauchinsky2010} which quote a sensitivity of $\sim$ 100~mV~cm$^{-1}$ with a spatial resolution of 7~$\mu$m. Recent experiments in strontium optical lattice clocks indicate that DC Stark shifts can lead to significant systematic errors \cite{Lodewyck2012}. Mapping the electric field at the $\sim$ 10~mV~cm$^{-1}$ level using Rydberg states could reduce this uncertainty by a factor of $\sim100$. We note that Rydberg excitation in the lattice clock has already been proposed as a sensitive black-body thermometer \cite{Ovsiannikov2011}. 


To create the initial cold atom ensemble, $^{88}$Sr atoms are cooled and trapped in a standard magneto-optical trap (MOT) operating on the 5s$^2$ $^1$S$_0$ $\rightarrow$ 5s5p $^1$P$_1$ transition at 461~nm ($\lambda_1$ in Fig.~\ref{fig:setup}~(a)), loaded from a Zeeman--slowed atomic beam.  A fluorescence image of the cloud is used to obtain the ground-state atom number and density distribution. Atom numbers of (5~$\pm$~0.5) $\times$ $10^6$ are achieved at densities of (2~$\pm$~0.2) $\times$ $10^9$~cm$^{-3}$ and temperatures of 5~$\pm$~0.5~mK.

The cold atoms are excited to a 5s$n$d $^1\mathrm{D}_2$ Rydberg state using a two-photon coherent population trapping (CPT, \cite{Arimondo1996,Schempp2010}) scheme ($\lambda_1$ and $\lambda_2$ in Fig.~\ref{fig:setup}~(a)).  The excitation lasers are counter-propagating to reduce the residual Doppler broadening and are pulsed on simultaneously for 1~$\mu$s.  A 3~$\pm$~0.2~G quantization field is applied, and both excitation beams are circularly polarized such that the $m_{\rm{J}} = + 2$ Rydberg state is excited.  The lasers are stabilized on resonance with their respective transitions using sub-Doppler frequency modulation spectroscopy $(\lambda_1$ \cite{Bjorklund1980}), and  electromagnetically-induced transparency ($\lambda_2$ \cite{Abel2009, Mauger2007}) in a dispenser-based vapour cell \cite{Bridge2009}.

The CPT spectrum is measured in the cold atoms by stepping the frequency of $\lambda_1$, with $\lambda_2$ held on resonance. For the 5s56d $^1$D$_2$ Rydberg state, we obtain a full-width at half maximum (FWHM) of 3.7~$\pm$~0.2~MHz in the low-intensity limit, compared to the 5s$^2$ $^1$S$_0$ $\rightarrow$ 5s5p $^1$P$_1$ transition width of 30.2~MHz.  Modelling of the lineshape using the optical Bloch equations (OBE) indicates that the spectral width is limited by the laser linewidths, which are approximately 1~MHz.  Beam $\lambda_1$ is collimated to a waist of 1.07~$\pm$~0.02~mm, while beam $\lambda_2$, is focused to a waist of 12.3~$\pm$~0.2~$\mu$m.  Hence, a narrow column of Rydberg atoms are created in the cold atom cloud, as seen in Fig.~\ref{fig:setup}~(b). 

To probe the spatial distribution of the Rydberg atoms, the second valence electron is excited using a laser tuned near the Sr$^{+} \mathrm{5s}_{1/2} \rightarrow \mathrm{5p}_{3/2}$ transition at $\lambda_3=408$\,nm (Fig.~\ref{fig:setup}~(a)). For low angular momentum Rydberg states, the doubly excited atom ionizes very rapidly ($\sim$~100~ps).  The excitation spectrum of the second electron is thus broadened to approximately 50~GHz  at $n\sim50$ \cite{Millen2010}. We drive this transition using an extended cavity diode laser stabilized to a wavemeter. The autoionizing laser $\lambda_3$ is focused to a waist of 10~$\mu$m and is pulsed on for 1~$\mu$s directly after the excitation beams have been extinguished to prevent direct photoionization.  The autoionizing laser is orthogonal to $\lambda_1$ and $\lambda_2$.  The ions that are produced are directed towards a micro-channel plate (MCP) via a weak electric field pulse (2.9~V~cm$^{-1}$) that is applied after the autoionizing beam is extinguished.  The electric field pulse is insufficient to field ionize the Rydberg atoms.

\begin{figure}[ht!]
  \includegraphics{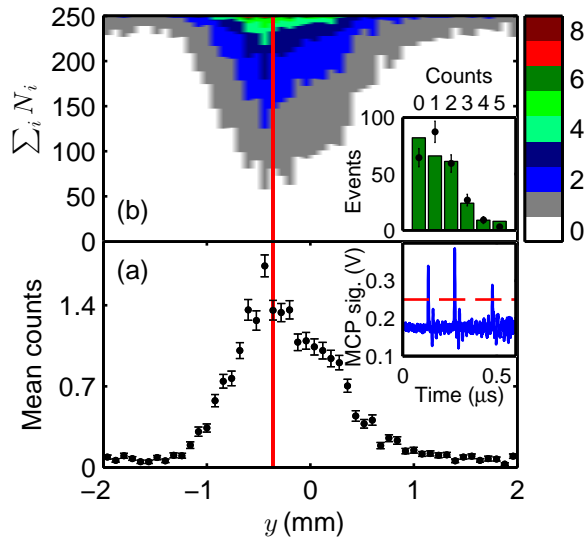}
  \caption{(Color online) (a) Mean ion signal as a function of $y$-axis position for a single slice. Error bars indicate the standard error.  Inset: Output signal from the MCP, showing single-ion pulses that are counted using a threshold (red line), which excludes noise and ringing. (b) The statistical distribution of counts corresponding to the slice shown, plotted as the sum of the number of repetitions $N_i$ in which $i$ counts were detected. Inset: The histogram of $N_i$ at the position indicated by the red line.The black dots with error bars show a Poisson distribution with the same mean count rate.}
  \label{fig:occurrences}
\end{figure}

An example of the ion signal generated by autoionizing the Rydberg atoms is shown in the inset to Fig.~\ref{fig:occurrences}~(a).  Individual ions arriving on the MCP generate a voltage pulse.  The individual voltage pulses are 2.5~ns wide and are distributed over an envelope of 2~$\mu$s due to differences in arrival time and stochastic processes within the MCP.  At low count rates, individual ion events can therefore be resolved and counted using a high-bandwidth digital oscilloscope. The ion count at each spatial position is collected for each of the 250 repetitions of the experiment.

The focused autoionizing laser is automatically scanned along the $y$-axis (long axis) of the ensemble by moving the focusing lens assembly using a high-precision translation stage. Using laser interferometry we measure an overall positional accuracy of 2~$\mu$m for a 1~mm displacement, which is significantly less than the laser spot size. By adjusting the angle of the scanning beam, the beam can also be moved in the $x$-direction, enabling us to build up a 2D-image composed of 1D-slices. The focal plane is imaged onto the CCD camera that is used for the fluorescence imaging, providing an additional check of the beam displacement, and a direct calibration of the position of the laser focus relative to the ground-state atom cloud. 

A 2D plot of the mean ion signal as a function of position is shown in Fig.~\ref{fig:setup}~(c). In the absence of background counts, the autoionization signal is proportional to the number of Rydberg atoms at the intersection of the excitation and autoionizing beams. The image in Fig.~\ref{fig:setup}~(c) therefore represents the density distribution of Rydberg atoms in the sample. Along the $y$-axis the Rydberg distribution is determined by the ground-state spatial distribution, as can be seen by the close agreement between the ion signal and a slice through the corresponding fluorescence image. Along the $x$-axis a much narrower distribution is measured (FWHM = 32~$\pm$~2~$\mu$m), arising from the tightly focused $\lambda_2$ beam.

A closer look at the mean ion count for a single slice is provided in Fig.~\ref{fig:occurrences}~(a).  The small, but non-zero, ion count rate in the extremities of the distribution is due to spontaneous ionization of the Rydberg atoms \cite{Robinson2000}.  Spontaneous ionization can occur anywhere in the Rydberg ensemble, unlike autoionization which only occurs at the intersection with the autoionization beam. Even so, at the centre of the ensemble the ratio of autoioniziation to spontaneous ionization is $\approx$ 20:1, showing the signal-to-noise ratio of the technique is very good, and justifying our previous disregard of the background signal. We have also verified that the density of background ions is far below that required for an ultracold plasma to form \cite{Millen2010,McQuillen2012}.

\begin{figure}[ht!]
	\includegraphics{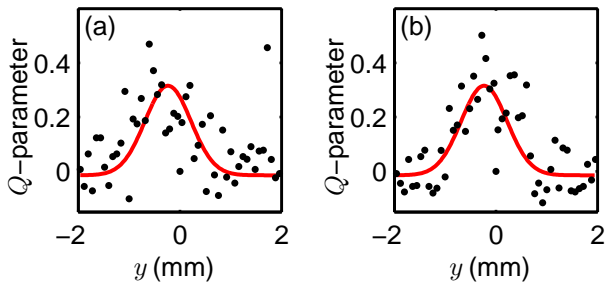}
	\caption{(Color online) (a) The Mandel $Q$-parameter as a function of $y$ for the data shown in Fig.~\ref{fig:occurrences}~(a). (b) Simulated Mandel $Q$-parameter, based on the same mean count rate and including technical noise.  The red lines indicate Gaussian fits to the corresponding mean ion signal.}
	\label{fig:Q_results}
\end{figure}

As well as the mean ion signal, this number-resolved detection method also provides spatially resolved measurements of the statistical distribution of counts. Figure.~\ref{fig:occurrences}~(b)  shows the complete counting distribution for the mean signal slice shown in Fig.~\ref{fig:occurrences}~(a). At each spatial sampling point we obtain the histogram of the number of counts, as shown in the inset. Access to the statistics, as well as to the mean is a major advantage of this technique since the onset of the quantum many-body Rydberg blockade regime \cite{Lukin2001} is associated with sub-Poissonian excitation statistics \cite{Liebisch2005}. From the statistical distribution shown in Fig.~\ref{fig:occurrences} we have calculated the Mandel $Q$-parameter, which characterizes the deviation from a Poisson distribution \cite{Mandel1979}. A plot of the spatially resolved $Q$-parameter is shown in Fig.~\ref{fig:Q_results}~(a). Except in the wings, the fluctuations are clearly super-Poissonian ($Q>0$), and the $Q$-parameter follows the shape of the mean ion count. The origin of these fluctuations is technical noise in our experiment; a simulation (Fig.~\ref{fig:Q_results}~(b)) that includes the independently-measured, shot-to-shot variation in atom number, laser detuning and intensity is in good agreement with experiment. Based on recent calculations of the interaction strength \cite{Vaillant2012}, we do not expect to observe the Rydberg blockade at the densities that can be reached in the relatively hot, high-scattering rate conditions of our magneto-optical trap. Indeed, we have seen no evidence of excitation suppression or sub-Poissonian statistics for Rydberg states up to $n=75$.

\begin{figure}[ht!]
	\includegraphics{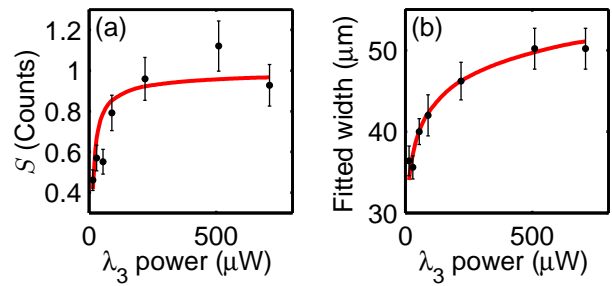}
	\caption{(Color online) (a) The fitted amplitude $S$ and (b) $1/e^2$ width of the Rydberg spatial distribution along the $x$-axis as a function of autoionizing laser power.  The red solid lines are the results of the OBE simulation described in the text.}
	\label{fig:408_width}
\end{figure}

The spatial resolution depends on both the achievable focusing of the autoionizing beam, and on its power. At a spot size of $\sim 10\,\mu$m, a few hundred microwatts is sufficient to saturate the autoionization probability for a 1~$\mu$s pulse. As the laser power is increased, this saturation leads to a broadening of the effective detector size, as autoionization becomes appreciable even in the wings of the laser beam. The saturation of the detection efficiency and the corresponding broadening are shown in Fig.~\ref{fig:408_width}. Using the OBE, and the known transition dipole matrix elements and decay rates, we have constructed a numerical model of the excitation and detection process that includes the finite beam sizes.  As shown in Fig.~\ref{fig:408_width} this model is in good agreement with the data. There are two fitting parameters: the overall detection efficiency, which we find to be 0.21~$\pm$~0.04; and the zero-power  width of the Rydberg distribution, which we find to be 32~$\pm$~2~$\mu$m. The latter is slightly broader than we would expect from the measured beam waists of the $\lambda_2$ and $\lambda_3$ lasers - possibly due to the laser beam mode quality after transmission through the vacuum chamber viewports. 

\begin{figure}
  \includegraphics{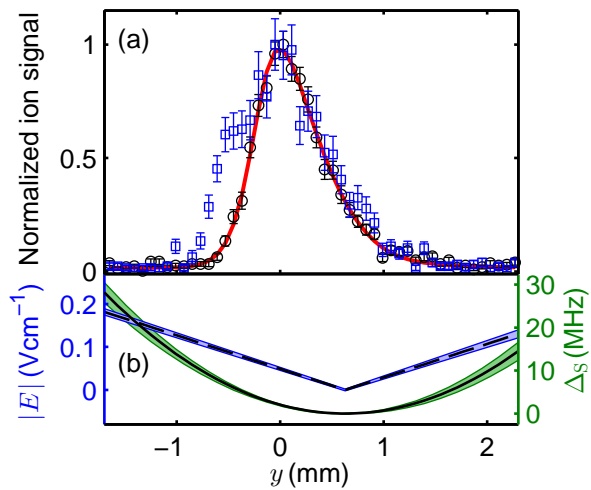}
  \caption{(Color online) (a) Slices through the Rydberg spatial distribution at $n=75$ with (black circles) and without (blue squares)  an externally applied inhomogeneous electric field.  The red solid line is the solution to the simulation described in the text.  (b) The dashed line shows the electric field extracted from the fit to the data above.  The solid line shows the Stark shift due to the electric field.  The colored bars show the standard deviations of the electric field and Stark shift.}
  \label{fig:electrometry}
\end{figure}

By exploiting the absence of interaction effects, and the large polarizability of Rydberg states, we can apply scanning autoionization microscopy to the spatially resolved measurement of electric fields. Figure~\ref{fig:electrometry} shows the effect of a spatially inhomogeneous electric field that is applied during the Rydberg excitation pulse on the Rydberg distribution. The wings of the cloud are Stark shifted out of resonance, and the Rydberg distribution and ground-state distribution no longer coincide. To extract the electric field from these measurements, we use the OBE model that fits to the data shown in Fig.~\ref{fig:electrometry}. The spatial variation of the electric field strength is described as a second-order polynomial with unknown, adjustable coefficients.  From this field profile, we calculate the position-dependent Stark shift using the polarizability of the 5s75d $^1$D$_2$ state \cite{Millen2011} and taking into account the angle between the applied electric and magnetic fields. The Rydberg excitation probability as a function of $y$ is then calculated using the OBE model, with the Stark shift included as a spatially varying detuning. By multiplying this excitation probability by the cloud shape obtained in the absence of electric fields, we generate a model Stark-shifted spatial distribution of the Rydberg excitations. The model distribution is  compared to the measured data as shown in Fig.~\ref{fig:electrometry}, and the coefficients of the polynomial are adjusted using a least-squares fitting routine. An additional fit parameter takes into account a possible additional (field-free) detuning. 

For the data shown in Fig.~\ref{fig:electrometry}~(a), the spatially dependent Stark shift $\Delta_{\rm{S}}$ and electric field magnitude $\vert$$E$$\vert$ is shown in Fig.~\ref{fig:electrometry}~(b). The quadratic term was found to be negligible, and we find an electric field gradient of  0.78~$\pm$~0.04~V~cm$^{-2}$ with an offset of $-0.049$~$\pm$~0.002~V~cm$^{-1}$.  The detuning of $\lambda_2$ from resonance was found to be $-2.2$~$\pm$~0.2~MHz. The error on these measurements includes the error from the fitting procedure, as well as an estimated 2.5~$\%$ error in the polarizability. Overall these results are in agreement with a simple model of the electric fields produced by the electrodes, but the precision of the measurements is significantly better than the accuracy of our calculations, which are limited to $\pm$~20~$\%$ due to uncertainties in the geometry and dielectric constants of the materials inside the vacuum.

In conclusion we have utilized the autoionizing process in strontium to map the spatial distribution of Rydberg atoms in a cold cloud of ground-state atoms.  We note that although a non-resonant focused laser beam could be used to photoionize any Rydberg atom, the ratio of the oscillator strength for the autoionizing transition relative to direct photoionization is $\approx$ 10$^{11}$ for the 5s56d $^1$D$_2$ state and scales as $n^6$ \cite{Gallagher1995}, hence using autoionization makes the experiment tractable.  Although this work focused on strontium the method is applicable to any multi-valence electron atom or molecule with autoionizing resonances. The method is sensitive enough to detect single Rydberg excitations, and provides a spatially resolved measurement of the excitation statistics.  In addition, the efficiency and resolution are independent of externally applied electromagnetic fields, enabling the high-sensitivity, high-resolution mapping of a weak, spatially inhomogeous electric field. The spatial resolution is set by the spot size of the autoionizing laser, and could be straightforwardly reduced to $1\,\mu$m using a higher numerical aperture, which is significantly smaller than the correlation length in strongly-interacting Rydberg gases \cite{Schausz2012}. In combination with the single-atom sensitivity and number resolving capability that we have demonstrated here, scanning autoionization microscopy therefore also provides an ideal tool for probing correlations in cold Rydberg gases.

We acknowledge useful discussions with E. Bridge and I. Hughes, and comments on the manuscript from K. Weatherill and J. Millen. This work was supported by EPSRC Grants no. EP/D070287/1 and EP/J007021/1.



\end{document}